\documentclass[aps,prl,preprint,groupedaddress]{revtex4-1}
\usepackage{graphicx}
\usepackage{amsmath}
\usepackage{siunitx}
\begin{document}
\title{Beyond the fundamental noise limit in coherent optical fiber links}
\author{C. E. Calosso$^{1}$, E. Bertacco$^{1}$, D. Calonico$^{1}$, C. Clivati$^{*,1}$, \\G. A. Costanzo$^{1,2}$, M. Frittelli$^{1}$, F. Levi$^{1}$, S. Micalizio$^1$, A. Mura$^{1}$, A. Godone$^{1}$}
\address{
$^{1}$Istituto Nazionale di Ricerca Metrologica INRIM, strada delle Cacce 91, 10135, Torino, Italy
\\
$^2$Politecnico di Torino, Corso Duca degli Abruzzi 24,10129, Torino, Italy \\
*Corresponding author: c.clivati@inrim.it
}

\begin{abstract}
It is well known that temperature variations and acoustic noise affect ultrastable frequency dissemination along optical fiber. Active stabilization techniques are in general adopted to compensate for the fiber-induced phase noise.  However, despite this compensation, the ultimate link performances remain limited by the so called delay-unsuppressed fiber noise that is related to the propagation delay of the light in the fiber.  In this paper, we demonstrate a data post-processing approach which enables us to overcome this limit. We implement a subtraction algorithm between the optical signal delivered at the remote link end and the round-trip signal. In this way, a \SI{6}{dB} improvement beyond the fundamental limit imposed by delay-unsuppressed noise is obtained. This result enhances the resolution of possible comparisons between remote optical clocks by a factor of 2. We confirm the theoretical prediction with experimental data obtained on a \SI{47}{km}  metropolitan  fiber link, and propose how to extend this method for frequency dissemination purposes as well.
\end{abstract}

\maketitle 

In the last decade, optical fiber links for frequency dissemination have become a key tool in frequency metrology. The frequency transfer through optical fibers \cite{pred,lopez2,williams,levi,sliv, fujieda, he,wang,marra} encompasses by more than four orders of magnitude the resolution of the present satellite techniques \cite{TWFT}, enabling a number of significant applications, such as the remote comparison of atomic clocks at the  \( 10^{-18} \) level of uncertainty \cite{chou,roseband}, and  the investigation of new frontiers in high resolution spectroscopy,   fundamental physics, geodesy \cite{chou2} and radioastronomy \cite{cliche}.  \\
Coherent frequency links through optical fiber are based on the transfer of an ultrastable laser frequency signal along a standard telecom fiber. 
It is well known that length variations due to temperature and mechanical stresses affect the fiber, resulting in a severe deterioration of the delivered signal phase stability. To overcome this difficulty, it is a common practice to adopt a Doppler noise cancellation scheme in which a double pass of the light in the fiber in opposite directions is exploited. Basically, from the link remote end, a part of the radiation is reflected back to the local laboratory, and here phase-compared to the transmitted optical radiation. Their beat note  allows the detection of the fiber phase noise, which is actively cancelled through a phase-locked loop. Hence, at the remote end, the beat note between the delivered laser and the local frequency reference is in principle not affected by the fiber phase noise \cite{williams}. \\
Actually, the round-trip fiber delay imposes an ultimate limit to the  link noise suppression. 
In fact, the phase noise  detection and cancellation is accomplished on the round-trip signal, whereas the delivered signal travels the fiber in a single pass and suffers from a residual noise due to the fiber delay. More specifically, even for a perfect Doppler cancellation at the local laboratory, the noise reduction factor of the delivered signal is  $\frac{(2 \pi f \tau)^2}{3}  $  for Fourier frequencies $f \ll \frac{1}{ \tau}$, where $\tau$ is the one-way propagation time along the link   \cite{williams}. \\
Although other techniques have been proposed as an alternative to Doppler-stabilized links in remote frequency comparisons \cite{calosso2vie}, it has been demonstrated that they are limited by the same amount of non-compensated noise.\\
In this letter, we describe a novel technique that improves the link performances  beyond this limitation and report on the experimental results we obtained.  The scheme we propose is still based on the typical architecture of Doppler-stabilized optical links. However, unlike the common procedure, 
 we only measure the phases of the local beat note and of the remote beat note, without implementing any active control loop. A convenient data processing allows to minimize the delay-unsuppressed noise, leading to a 6 dB improvement of the fundamental limit. \\
The beat notes are synchronously measured with digital phasemeters \cite{calosso1}. The digital implementation ensures a precise timing of the measurements, which is essential for the synchronous acquisition, is reliable and easily  adapted to perform more complex tasks. This technique can be in fact also extended to simultaneously implement  an active system for frequency dissemination purposes.\\
We tested our system on a \SI{47}{km}-long optical link based on a Dense Wavelength Division Multiplexed (DWDM) architecture and buried in the metropolitan area of Turin (Italy), with both ends in our laboratory at the Italian National Metrology Institute (INRIM).\\
As in the Doppler cancellation technique, 
we intend to compensate the phase of the optical signal delivered at the remote link end, $\varphi_\text{r}(t)$, by detecting the optical phase on the round-trip signal, $\varphi_\text{rt}(t)$. \\
As a first step, we evaluate the relation between the fiber phase noise and the transmitted signal phase noise in terms of their time evolution. \\
If $\delta\varphi(z,  t) dz$ is the phase noise on the fiber at time $t$ and position $z$, $\varphi_\text{r}(t)$ can be written as:
\begin{equation}
\label{eq:1}
\varphi_\text{r}(t)= \int_0^L \delta \varphi \left(z, t-\tau+\frac{z}{c_{n}}\right)dz 
\end{equation}
where $L$ is the link length and $c_n$ is the speed of light in the fiber. The phase noise $\varphi_\text{rt}(t)$ accumulated by the light on the round-trip travel is given by:
\begin{equation}
\begin{split}
\label{eq:2}
\varphi_\text{rt}(t)&= \int_0^L \left[\delta \varphi\left(z,t-2\tau+\frac{z}{c_{n}}\right) + \delta \varphi\left(z, t-\frac{z}{c_{n}}\right) \right] dz \\
&\approx \int_0^L 2\delta \varphi\left(z,t-\tau\right)dz
\end{split}
\end{equation}
In Eq. (\ref{eq:2}), the approximation assumes a linear evolution of the fiber perturbation at position $z$ between the forward and the backward paths. This is justified for perturbations which act on timescales much longer than $\tau$, as in the case of interest in ultrastable optical links applications.\\
Within this approximation,  the most intuitive approach to compensate the phase noise in $z=L$ consists of subtracting half of the round-trip signal phase from the forward signal phase:
\begin{equation}
\label{eq:3}
\varphi_\text{r,comp}(t) =   \varphi_\text{r} \left(t\right)-\frac{1}{2} \varphi_\text{rt}(t)  
\end{equation} 
where the factor $1/2$ takes into account that the noise estimation is performed on a double pass in the fiber. 
 From Eqs. (\ref{eq:1}-\ref{eq:3}), 
 $\varphi_\text{r,comp}(t)$ can be calculated as:
\begin{equation}
\begin{split}
\label{eq:4}
\varphi_\text{r,comp}(t) & \approx  \int_0^L \left[\delta \varphi\left(z,t-\tau+\frac{z}{c_{n}}\right) - \delta \varphi\Big(z, t-\tau\Big) \right] dz \\
& \approx  \int_0^L \frac{z}{c_{n}}\frac{\partial }{\partial t} \delta \varphi (z, t) dz \\
& = \int_0^L h(z, t)  \delta \varphi(z, t) dz
\end{split}
\end{equation} 
In the second line of Eq. (\ref{eq:4}), the difference between $\delta\varphi(z, t-\tau +z/c_{n})$ and $\delta\varphi(z, \ t-\tau)$ has been rewritten in terms of their time derivative. Again, this is justified for perturbations which act on timescales much longer than $\tau$. In the last step, we introduced the function  $h(z, \ t) =\frac{z}{c_{n}}\frac{\partial }{\partial t}$ as the impulse response that, for each $z$, processes the local phase perturbation. 
Considering the contribution of each fiber segment with length $dz$, we can apply the fundamental theorem of spectral analysis \cite{papoulis} that expresses the power spectrum of the output of a  linear and time-invariant system in terms of the input power spectrum. In our case,
\begin{equation}
\label{eq:5a}
S_\text{r,comp} (z,f)=\vert H(z,f) \vert^2 S_{\delta \varphi} (z,f) 
\end{equation}
where $S_\text{r,comp}(z,f)$ and  $S_{\delta \varphi} (z,f)$ are the noise spectral density contributions of a section $dz$ of fiber at position $z$ for the compensated and free fiber respectively. $\vert H(z,f) \vert^2=\left(2\pi f \frac{z}{c_{n}} \right)^2$ is the square modulus of the $h(z, \ t)$ Fourier transform, i.e. the transfer function. If the noise is uncorrelated along the fiber, the noise contributions coming from different  $z$  positions are independent and sum up, 
  so that we can write:
\begin{equation}
\label{eq:5}
S_\text{r,comp} (f)=\int_0^L \vert H(z,f) \vert^2 S_{\delta \varphi} (z,f) dz.
\end{equation}
Assuming, in addition, that the fiber noise is uniformly distributed along the link, i.e. $S_{\delta \varphi} (z,f)= S_\text{r} (f)/L $, the integration leads to:
\begin{equation}
\label{eq:6}
S_\text{r,comp} (f) =\frac{1}{3} (2 \pi f \tau )^2 S_\text{r} (f)
\end{equation}
where $S_\text{r} (f)$ is the phase noise power spectrum of the fiber. Although this result has been obtained in a passive approach, where the round-trip fiber noise is measured and used to correct the forward signal, it is interesting to note that the same limitation is found in the case of actively Doppler-compensated and two-way links \cite{williams,calosso2vie}. We point out that we do not assume that the power spectrum of $\varphi(t)$ is the square modulus of its Fourier transforms.  
In our approach, Eq. (\ref{eq:6}) has been deduced using the general definition of power spectral density in terms of the Fourier transform of its autocorrelation, which is the formal definition in the case of  random processes \cite{papoulis}.\\
In order to improve the limitation imposed by Eq. (\ref{eq:6}), we reconsider Eq. (\ref{eq:3}) supposing that it is possible to perform the subtraction between arbitrarily time-delayed samples, with delay $\alpha$:
\begin{equation}
\label{eq:7}
\tilde{\varphi}_\text{r,comp}(t) =   \varphi_\text{r} \left(t \right)-\frac{1}{2} \varphi_\text{rt} (t+\alpha) 
\end{equation}
where we indicated as $\tilde{\varphi}_\text{r,comp}$ the corresponding compensated phase. With the same assumptions adopted previously, i.e. noise uncorrelated and uniformly distributed along the link, Eqs. (\ref{eq:5}) and (\ref{eq:6}) are now modified into:
\begin{equation}
\label{eq:8}
\begin{split}
\tilde{S}_\text{r,comp} (f)=&\int_0^L \vert \tilde{H}(z,f) \vert^2 S_{\delta \varphi} (z,f) dz\\
=& \frac{1}{3}(2 \pi f)^{2}\left(\tau^2-3\alpha \tau+3\alpha^{2}\right)S_\text{r} (f)
\end{split}
\end{equation}
where $\vert \tilde{H}(z,f) \vert^2=(2\pi f)^2\left(\frac{z}{c_{n}}-\alpha\right)^2$. It can be seen that the noise conversion is minimized for $\alpha= \tau/2$ leading to:
\begin{equation}
\label{eq:9}
\tilde{S}_\text{r,comp} (f) =\frac{1}{12} (2 \pi f \tau )^2 S_\text{r} (f)
\end{equation}
The remarkable result expressed by Eq. (\ref{eq:9}) is that performing the subtraction of time-shifted phase data, the fundamental limit shown in Eq. (\ref{eq:6}) is overcome by 6 dB. More specifically, the noise compensation is optimized by anticipating the round-trip signal of $\tau/2$. This of course can be done only by post-processing the phases $\varphi_\text{r}$ and $\varphi_\text{rt}$ acquired at both sides of the fiber.\\
To verify the theoretical prediction of Eq. (\ref{eq:9}), we implemented an optical link using \SI{47}{km} of fiber placed in the metropolitan area of Torino. Accordingly, $\tau = $ \SI{235}{\micro\second} for this loop.\\
The link is realized on a DWDM architecture and is shared with other users: channel 44 of the International Telecommunication Union grid is dedicated to this experiment, whereas channels 21 and 22 are occupied by the Internet traffic. \\
\begin{figure}
\centering
\includegraphics[width=1\columnwidth]{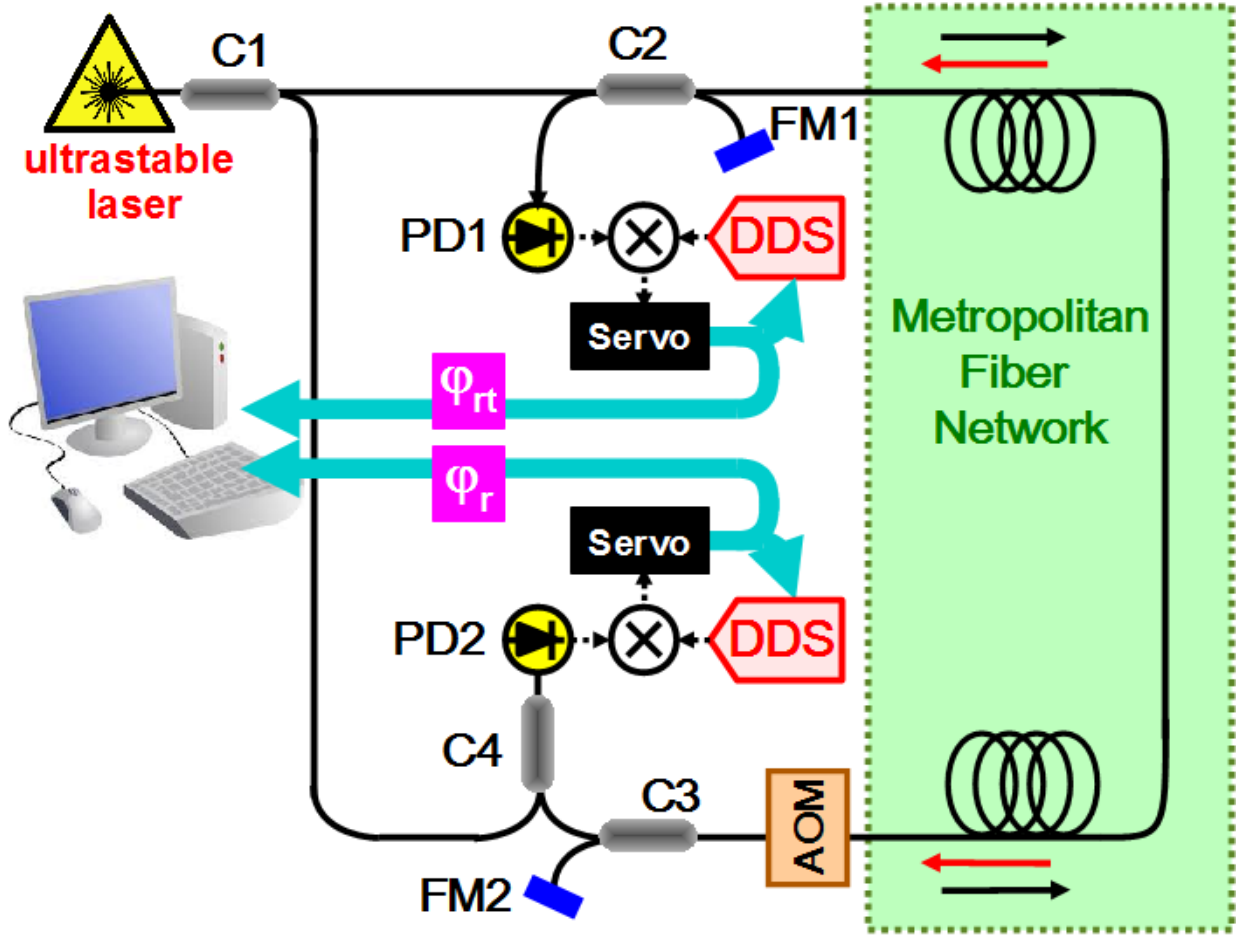}
\caption{(Color online) The setup of the optical link. The ultrastable laser is splitted into two beams. A part is used as a local oscillator; the remainder travels the link and, at the remote end, is frequency shifted by the acousto-optic modulator AOM, extracted, and compared to the local oscillator on photodiode PD2. A portion of the light is reflected by the Faraday mirror FM2 and travels the link in the backward direction; the round-trip light is compared to the local oscillator on photodiode PD1; C1-C4 represent optical couplers. The beat notes on PD1 and PD2 are tracked by two Direct Digital Synthesizers (DDSs); the phase corrections $\varphi_\text{rt}$ and $\varphi_\text{r}$ are also sent to a PC for measurement. \label{fig:setup}}
\label{fig:setup}
\end{figure}
The experimental setup is described in Fig. \ref{fig:setup}. The optical source is a fiber laser frequency-stabilized on a high-finesse Fabry-P\'{e}rot cavity at the $10^{-14}$  stability level at \SI{1}{s} \cite{clivati}. It is split into two parts: a small fraction of the optical power is used as a local oscillator, the remainder travels the optical link.  At the remote link end, the radiation is frequency shifted by \SI{40}{MHz} by the acousto-optic modulator AOM to distinguish the round-trip signal from the stray reflections along the fiber. A fraction of the delivered radiation is extracted and compared to the original one on photodiode PD2; this enables to directly measure the link performances, neglecting the laser noise contribution. The remainder is reflected by a Faraday mirror (FM2) to the local laboratory, and here compared to the original laser on photodiode PD1. \\
The phases of the beat notes on PD1 and PD2 represent the phase noises of the round-trip and of the forward signal respectively. They are measured with a digital and synchronous electronic system in which Two Direct Digital Synthesizers (DDSs) \cite{calosso1} track the beat notes through a phase-locked loop (PLL). In particular, a double balanced mixer is used as phase discriminator and a Field Programmable Gate Array (FPGA) implements the servo and drives the phase of the DDS. Within the PLL bandwidth, the data sent to the DDSs coincide with the phases of the beat notes numerically expressed, which can thus be retrieved.\\
In our experiment, we record $\varphi_\text{r}$ and $\varphi_\text{rt}$ with a sampling rate of \SI{4}{kHz} and a measurement bandwidth of \SI{2}{kHz}. Then, we implement the algorithms exposed in Eq. (\ref{eq:3}) and Eq. (\ref{eq:7}) by post processing.\\
\begin{figure}
\centering
\includegraphics[width=1\columnwidth]{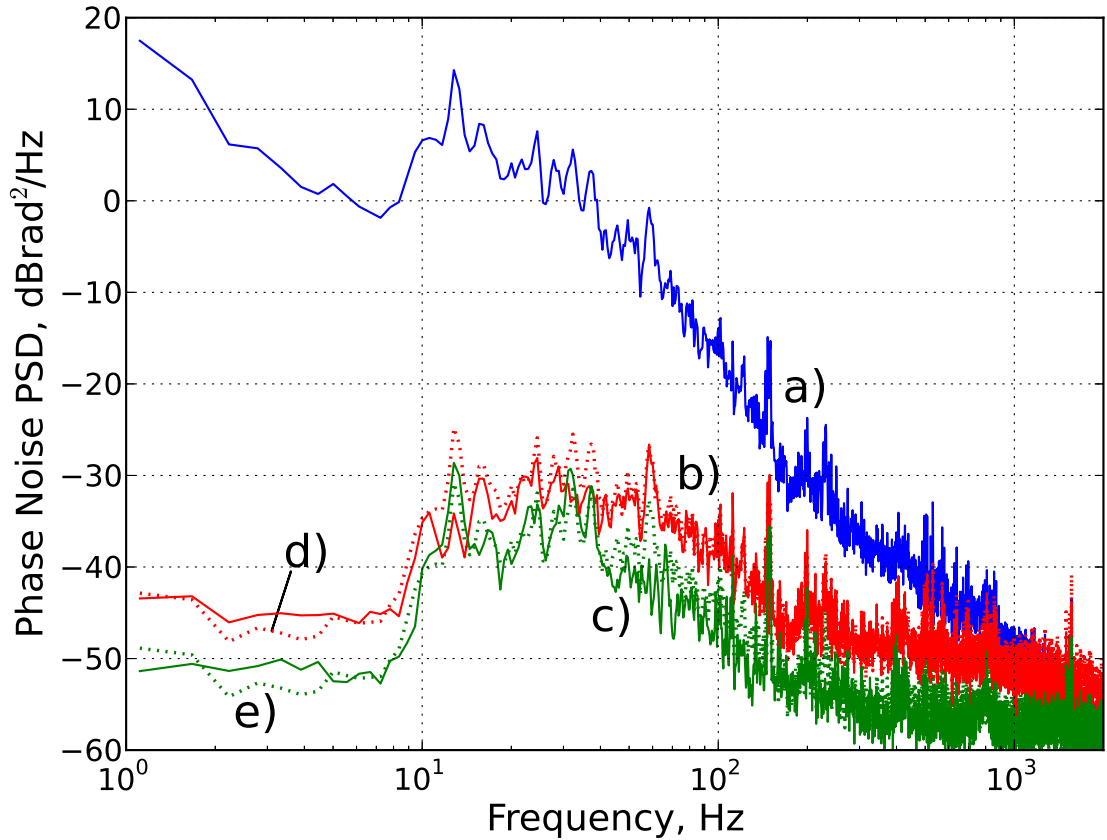}
\caption{(Color online) The phase noise power spectra measured at the remote end. Blue line (a): without compensation, red line (b): the compensated phase noise with synchronous subtraction, green line (c): the compensated phase noise when the subtraction is performed with a $\tau/2$ shift between samples. Dotted lines (d) and (e) represent the prediction in the two cases, according to Eq. (\ref{eq:3}) and Eq. (\ref{eq:7}) respectively. \label{fig:measure}}
\end{figure}
Figure \ref{fig:measure} shows the power spectrum densities of the signals of interest: the transmitted radiation without compensation  $S_\text{r}(f)$ (blue line, a) and with compensation, both when the samples are subtracted synchronously (red line, b) and when a time shift of $\tau/2$ is applied (green line, c). Dotted lines (d) and (e) represent the expected limitations in the two cases. It can be seen that the measurements are in good agreement with the predictions, and that the noise suppression is improved by \SI{6}{dB} for frequencies $f \ll \frac{1}{\tau}$ when the samples are time-shifted by $\tau/2$.
These results demonstrate that the comparison of distant clocks can be improved by a factor of 2 in terms of frequency stability if a proper subtraction algorithm is implemented; this increases the resolution of the comparison, or, which is the same, allows the reduction of the measurement time needed for the comparison.\\ 
It is interesting to note that this scheme can be applied to existing Doppler-stabilized links as well.  To figure out, in principle, how this scheme can be implemented, we consider again the setup of Fig. \ref{fig:setup}, where a second acousto-optic modulator AOM2 is inserted  between the coupler C2 and the fiber as an actuator for the active link stabilization. We consider the effect of its additional phase modulation $\varphi_\text c$ on the the round-trip and on the forward signals. Accordingly, we obtain:
\begin{equation}
\label{eq:10}
\begin{split}
\varphi_\text{r}(t)&= \varphi^\text{cl}_\text{r}(t) - \varphi_\text c(t-\tau)\\
\varphi_\text{rt}(t)&= \varphi^\text{cl}_\text{rt}(t) - \varphi_\text c(t-2\tau) - \varphi_\text c(t)\\
\end{split}
\end{equation}
The superscript ``cl'' identifies the relevant signals in the closed loop approach. According to Eq. (\ref{eq:10}), we can then rewrite Eq. (\ref{eq:7}) as a function of the signals in closed loop condition:
\begin{equation}
\label{eq:epswp_cl}
\begin{split}
\tilde{\varphi}_\text{r,comp}(t)&= \varphi_\text{r}(t)-\frac{1}{2} \varphi_\text{rt}\Big(t+\frac{\tau}{2}\Big) \\
					 &= \varphi^\text{cl}_\text{r}(t) - u(t)
\end{split}
\end{equation}
where we define $u(t)$ as follows
\begin{equation}
\label{eq:upgrade}
u(t) = \varphi_\text c(t-\tau) + \frac{\varphi^\text{cl}_\text{rt}(t+\frac{\tau}{2}) - \varphi_\text c(t-\frac{3}{2}\tau) - \varphi_\text c(t+\frac{\tau}{2})}{2}
\end{equation}
From a practical point of view, $u(t)$ represents an upgrade signal, which can be measured in the local laboratory by synchronously acquiring the phase-correction $\varphi_c(t)$ imposed to AOM2, and the PD1 beat note phase when the controller is active. The upgrade signal can then be applied to the phase measurement at the remote end, to improve the comparison during post-processing.\\
In this way, the optical frequency dissemination  can be performed at the classical limit; in addition, the phase-measurement can be further processed off line to obtain the 6 dB improvement. \\
In conclusion, we demonstrated from a theoretical and an experimental point of view that the fundamental limit in the phase-stabilization of coherent optical links can be overcome by time-shifting the phase correction applied to the forward signal. 
Without a loss of generality, this improvement was demonstrated in a passive approach, where there is no active cancellation of the fiber noise. 
Furthermore, we have also suggested how to upgrade the apparatus to perform two tasks at the same time: the delivery of an ultrastable frequency signal and the off-line  processing  to increase the resolution in a frequency comparison. \\

We thank Luca Lorini for useful discussion and careful reading of the manuscript, and Consortium GARR for technical help with the fibers.\\
This  work  was  partly funded  by  Compagnia  di  San  Paolo,  by MIUR under Progetto Premiale 2012, and  by the  EMRP  programs SIB02-NEAT-FT and IND55-MClocks.  The  EMRP  is  jointly  funded  by  the  EMRP  participating countries within EURAMET and the European Union.\\

\end{document}